\documentclass[10pt,conference]{IEEEtran}

\usepackage{cite}
\usepackage{threeparttable}

\ifCLASSINFOpdf
   \usepackage[pdftex]{graphicx}
   \graphicspath{{fig/}}
   \DeclareGraphicsExtensions{.pdf,.jpeg,.png}
\else
   \usepackage[dvips]{graphicx}
   \graphicspath{{fig/}}
   \DeclareGraphicsExtensions{.eps}
\fi

\usepackage[cmex10]{amsmath}
\usepackage{array}
\usepackage{multirow}

\usepackage{amsfonts}

\ifCLASSOPTIONcompsoc
  \usepackage[caption=false,font=normalsize,labelfont=sf,textfont=sf]{subfig}
\else
  \usepackage[caption=false,font=footnotesize]{subfig}
\fi

\usepackage{url}

\renewcommand{\baselinestretch}{1}

\IEEEoverridecommandlockouts

\begin{document}

\title{An Improved DOA Estimation Method for a Mixture of Circular and Non-Circular Signals Based on Sparse Arrays}

\author{Jingjing~Cai,
        Wei~Liu, 
        Ru~Zong,
        and Yangyang~Dong
\thanks{This work was supported by the National Natural Science Foundation of China (No.61805189 and No.61901332) and Natural Science Basic Research Plan in Shaanxi Provincial of China (No.2018JQ6068).}
\thanks{Jingjing Cai, Ru Zong and Yangyang Dong are with the Department of Electronic and Engineering,
Xidian University, Xi'an, 710071, China (e-mail: cjjyoyo@126.com,zongru@xidian.edu.cn,dongyangyang2104@126.com).}
\thanks{Wei Liu is with the Department of Electronic and Electrical Engineering,
University of Sheffield, Sheffield, S1 3JD, UK (e-mail: w.liu@sheffield.ac.uk).}
}
\maketitle

\begin{abstract}
Sparse arrays have attracted a lot of interests recently for their capability of providing more degrees of freedom than traditional uniform linear arrays. For a mixture of circular and noncircular signals, most of the existing direction of arrival (DOA) estimation methods are based on various uniform arrays. Recently, a class of DOA estimation algorithms based on sparse arrays was developed for a mixture of circular and noncircular signals. To further improve its performance, in this work, a modified algorithm is presented, which can resolve the same number of signals, and simulation results are provided to verified its performance.
\end{abstract}

\IEEEpeerreviewmaketitle

\section{Introduction}
\label{Introduction}
In array signal processing, direction of arrival (DOA) estimation is a very important research area and many existing algorithms usually assume either explicitly or implicitly that the incoming signals are circular, but in practice there are many noncircular signals, such as pulse amplitude modulation (PAM) and binary phase shift keying (BPSK) modulated signals\cite{steinwandt15a}. In the past, a lot of efforts have been made to explicitly exploit
the noncircularity features of the impinging array signals to improve the performance in parameters estimation\cite{charge01,abeida06,liu08l,abeida08,huang2010performance,steinwandt14,liu16h,gao08a,steinwandt15a,steinwandt16a,steinwandt16b}, where the main approach is to extend the array aperture effectively by augmenting the received array signals with their conjugate version, with the dimension of the new covariance matrix doubled.

 On the other hand, sparse arrays have attracted more and more attention and become a very important research area due to their ability to provide much more degrees of freedom (DOFs) than traditional uniform linear arrays (ULAs), and two of the representative sparse array structures are the co-prime arrays (CPAs)\cite{ppal11a,vaidyanathan11a,zhang13a} and nested arrays (NAs) \cite{ppal10a,shen15a,liu16aa,liu16bb,liu17a}. Many methods have been proposed for DOA estimation based on such arrays, such as \cite{pal2011coprime,liu2015Remarks,vaidyanathan11a,liu15e,qin15a,cai16,shi17a,cai17,cai17b,cai18b,cai18a}. However, almost all these algorithms are focused on the circular signal situations, except for the work in \cite{cai17b,cai18a} which considered the situation with a mixture of circular and noncircular signals.

 In \cite{cai18a}, two types of estimation algorithms are proposed. One is based on the subspace method while the other is based on the sparse representation principle. For the subspace based method, two algorithms are derived, called unequal length (UL) and unequal length plus (ULP), respectively, which use polynomial rooting in the estimator to find the solution. It adopts a general signal model by considering the mixture signals as a whole and does not differentiate the circular and noncircular signals in the formulation. Although the performance of the UL and ULP algorithms is not good for low SNR circumstances, compared to the sparse representation based algorithm, it has a much lower computational complexity. In order to improve the performance of the subspace based algorithms, a new algorithm called improved MUSIC (I-MUSIC) algorithm is proposed in this work based on a new formulation. The number of signals that can be resolved by the proposed I-MUSIC algorithm is analysed, which turns out to be the same as the UL algorithm. As demonstrated by computer simulations, overall a much better performance has been achieved by the proposed method.

This paper is organized as follows. The signal model for the mixed signals based on the sparse array is presented in Sec. \ref{sec:data model}, and the I-MUSIC algorithm is proposed in Sec. \ref{sec:ULP}.  Simulation results are provided in Sec. \ref{sec:simulation}, and conclusions are drawn in Sec. \ref{sec:conclusion}.
\section{sparse array signal model}
\label{sec:data model}
Consider an $M$-sensor linear sparse array with its sensors located at $\{p_1\cdot d, p_2\cdot d,\dots,p_M\cdot d\}$, where $d$ is the unit inter-element spacing and $p_1, p_2,\dots, p_M$ are integers. There are $K$ far-filed stationary and uncorrelated narrowband sources impinging on the array from directions $\{\theta_1,\theta_2,\dots, \theta_K\}$. Assume the signals are a mixture of circular and non-circular signals, and the first $K_{nc}$ signals are non-circular, while the last $K_c$ signals are circular, such that $K_{nc}+K_c=K$. The received signal vector $\mathbf{x}(t)$ of the array can be given by
\begin{equation}
\begin{split}
        \mathbf{x}(t)&= \sum_{k=1}^{K_{nc}} \mathbf{a}(\theta_k)s_{k}(t)+\sum_{k=K_{nc}+1}^K \mathbf{a}(\theta_k)s_{k}(t)+\mathbf{n}(t)  \\
       & =\mathbf{A}\mathbf{s}(t)+\mathbf{n}(t)\;\;\;\;t=1,2,...,N
\end{split}
\end{equation}
where $s_{k}(t)$ represents the $k$th source signal, and its first $K_{nc}$ sources are noncircular while the last $K_c$ are circular, $N$ is the total number of snapshots, $\mathbf{n}(t)$ is the uncorrelated noise with zero-mean and covariance matrix $\sigma ^2 \mathbf{I}$, $\mathbf{a}(\theta_k)$ is the steering vector corresponding to the $k$th signal, and $\mathbf{A}$ is the steering matrix, given by
 \begin{equation}
  \begin{split}
  &\mathbf{a}(\theta_k)=[e^{-j(2\pi p_1dsin\theta_k/\lambda)},\dots, e^{-j(2\pi p_Mdsin\theta_k/\lambda)}]^T \\
  &\mathbf{A}=[\mathbf{a}(\theta_1),\mathbf{a}(\theta_2),\dots,\mathbf{a}(\theta_K)]
  \end{split}
 \end{equation}
with $(\cdot)^T$ denoting the transpose operation.

The covariance matrix $\mathbf{R}_{xx}$ and pseudo covariance matrix $\mathbf{R}_{xx^\ast}$ of the received array signal are respectively expressed as
\begin{equation}
\begin{split}
   &\mathbf{R}_{xx}=E\{\mathbf{x}(t)\mathbf{x}^H(t)\}=\sum_{k=1}^{K}\eta_k\mathbf{a}(\theta_k)\mathbf{a}^H(\theta_k)+\delta^2\mathbf{I} \\
   &  =\mathbf{A} \mathbf{R}_s \mathbf{A}^H+\delta^2\mathbf{I}\\
    &\mathbf{R}_{xx^\ast}=E\{\mathbf{x}(t)\mathbf{x}^T(t)\}=\sum_{k=1}^{K}\rho_k e^{j\psi_k}\eta_k\mathbf{a}(\theta_k)\mathbf{a}^T(\theta_k)\\
     &=(\mathbf{A}^*\mathbf{V}^*) \mathbf{R}_s (\mathbf{A}^*\mathbf{V}^*)^H
\end{split}
\end{equation}
where  $(\cdot)^H$ denotes the Hermitian transpose operation, $(\cdot)^*$ represents the conjugate operation, $\eta_k$ is the power of the $k$th source, $\rho_k$ and $\psi_k$ are the noncircularity rate and phase of the $k$th noncircular signal. For circular signals, $\rho_k=0$, and it is $0<\rho_k\leq 1$ for noncircular signals. For strictly noncircular signals, such as BPSK modulated signals, we have $\rho_k=1$. The source covariance matrix $\mathbf{R}_s$ and the $K\times K$ matrix $\mathbf{V}$ are defined as
\begin{equation}
\begin{split}
    &\mathbf{R}_s=diag \{\eta_1,\eta_2,\cdots,\eta_K\}\\
    &\mathbf{V}=diag \{\rho_1 e^{j\psi_1},\cdots,\rho_{K_{nc}}e^{j\psi_{K_{nc}}},0,\cdots,0\}
\end{split}
\end{equation}
with $diag\{\cdot\}$ represents the corresponding diagonal matrix. It can be seen that $\mathbf{R}_s$ is of full rank as the received signals are uncorrelated; if all the sources are circular, $\mathbf{R}_{xx^*}$ will vanish, and if the signals are a mixture of circular and noncircular signals, both $\mathbf{R}_{xx}$ and $\mathbf{R}_{xx^\ast}$ are non-zero valued.

In practice, with the finite number of snapshots, the covariance and pseudo covariance matrices can be estimated as follows
\begin{equation}
\begin{split}
    &\mathbf{\hat R}_{xx}=(1/N)\sum_{t=1}^N\mathbf{x}(t)\mathbf{x}^H(t)\\
    &\mathbf{\hat R}_{xx^{\ast}}=(1/N)\sum_{t=1}^N\mathbf{x}(t)\mathbf{x}^T(t)
\end{split}
 \end{equation}
\section{The Proposed Algorithm}
\label{sec:ULP}
 For the subspace based DOA estimation algorithms, the first step is to construct an extended covariance matrix with a much larger dimension based on the covariance matrix and pseudo covariance matrix. In order to do that, we firstly vectorize the matrices $\mathbf{R}_{xx}$ and $\mathbf{R}_{xx^*}$ as $\mathbf{r}_{xx}$ and $\mathbf{r}_{xx^*}$, which can be written as
 \begin{equation}
 \begin{split}
  \mathbf{r}_{xx}&=\sum_{k=1}^{K}\eta_k\mathbf{a}^*(\theta_k)\otimes \mathbf{a}(\theta_k)+[\delta^2,\dots,\delta^2]^T  \\
  &=\mathbf{A}_{xx}\mathbf{v}_{xx}+[\delta^2,\dots,\delta^2]^T\\
     \mathbf{r}_{xx^{\ast}}&=\sum_{k=1}^{K_{nc}}\rho_k e^{j\psi_k}\eta_k\mathbf{a}(\theta_k)\otimes \mathbf{a}(\theta_k)\\
   &  =\mathbf{A}_{xx^{\ast}}\mathbf{v}_{xx^{\ast}}\\
 \end{split}
 \end{equation}
 with
\begin{equation}
\begin{split}
   &  \mathbf{A}_{xx}=[\mathbf{a}^*(\theta_1)\otimes \mathbf{a}(\theta_1),\dots,\mathbf{a}^*(\theta_K)\otimes \mathbf{a}(\theta_K)] \\
 &\mathbf{A}_{xx^{\ast}}=[\mathbf{a}(\theta_1)\otimes \mathbf{a}(\theta_1),\dots,\mathbf{a}(\theta_K)\otimes \mathbf{a}(\theta_K)]\\
 & \mathbf{v}_{xx}=[\eta_1,\dots,\eta_K]^T \\
 &\mathbf{v}_{xx^{\ast}}=[\rho_1e^{j\psi_1}\eta_1,\dots,\rho_{K_{nc}}e^{j\psi_{K_{nc}}}\eta_{K_{nc}},0,\dots,0]^T\\
 \end{split}
\end{equation}
where $\otimes$ denotes the Kronecker product. Clearly, the elements of $\mathbf{r}_{xx}$ and $\mathbf{r}_{xx^*}$ are related to $(p_i-p_j)$ and $(p_u+p_v)$, separately, where $p_i$,$p_j$,$p_u$ and $p_v$ are arbitrary values chosen from $p_1,\dots,p_M$.

Then, we order the elements of $\mathbf{r}_{xx}$ and $\mathbf{r}_{xx^{\ast}}$ with an increasing steering vector index, separately. If some elements have the same steering vector index, the mean value of these elements is used. Moreover, non-consecutive elements are removed from the ordered vectors. Using this strategy, we can finally obtain two new vectors $\mathbf{r}_{d}$ and $\mathbf{r}_{s}$. Assuming that the length of these two new vectors are $L_1$ and $L_2$, the two vectors can be written as
\begin{equation}\label{eq:rdrs}
\begin{split}
   &  \mathbf{r}_d=\sum_{k=1}^{K}\eta_k\mathbf{a}_d(\theta_k)+[\delta^2,\dots,\delta^2]^T \\
   &  \mathbf{r}_s=\sum_{k=1}^{K}\rho_k e^{j\psi_k}\eta_k\mathbf{a}_s(\theta_k)\\
\end{split}
\end{equation}
with
\begin{equation}\label{eq:ads}
  \begin{split}
  & \mathbf{a}_d(\theta_k)=[e^{-j(2\pi \frac{L_1-1}{2}dsin\theta_k/\lambda)},\dots, e^{j(2\pi \frac{L_1-1}{2} dsin\theta_k/\lambda)}]^T \\
  & \mathbf{a}_s(\theta_k)=[e^{-j(2\pi \Delta p d sin\theta_k/\lambda)},\dots, e^{-j(2\pi (\Delta p+L_2-1) dsin\theta_k/\lambda)}]^T
  \end{split}
 \end{equation}
where $\Delta p$ is the first number produced in the $L_2$-length consecutive sequence. It is clear that $\mathbf{a}_d(\theta_k)$ is corresponding to $\mathbf{a}(\theta_k)\mathbf{a}^H(\theta_k)$, while $\mathbf{a}_s(\theta_k)$ is corresponding to $\mathbf{a}(\theta_k)\mathbf{a}^T(\theta_k)$. In another word, the elements of $\mathbf{a}_d(\theta_k)$ related to $(p_i-p_j)$, while $\mathbf{a}_s(\theta_k)$ related to  $(p_u+p_v)$.

Based on $\mathbf{r}_{d}$ and $\mathbf{r}_{s}$, we construct an extended covariance matrix $\mathbf{R}_u$, which is written like this
\begin{equation}\label{eq:Ru}
\begin{split}
  \mathbf{R}_{u} =\left[ \begin{array} {cc} \mathbf{R}_1&\mathbf{R}_2\\\mathbf{R}_2^H&\mathbf{R}_3 \end{array} \right]
\end{split}
\end{equation}
where
\begin{equation}
\begin{split}
  &\mathbf{R}_1= \left[ \begin {array}{ccc} \mathbf{r}_d(\frac{L_1+1}{2})&\dots& \mathbf{r}_d(L_1)\\
  \vdots&\ddots&\vdots\\ \mathbf{r}_d(1)&\dots&\mathbf{r}_d(\frac{L_1+1}{2}) \end{array}\!\right],\\
  &\mathbf{R}_2= \left[ \begin {array}{ccc} \mathbf{r}_s(1)&\dots& \mathbf{r}_s(L_2-\frac{L_1-1}{2}),\\
  \vdots&\ddots&\vdots\\ \mathbf{r}_s(\frac{L_1+1}{2})&\dots&\mathbf{r}_s(L_2) \end{array}\!\right]\\
  &\mathbf{R}_3=\left[ \begin {array}{ccc} \mathbf{r}_d(\frac{L_1+1}{2})&\dots& \mathbf{r}_d(L_2-\frac{L_1-1}{2})\\ \vdots&\ddots&\vdots\\ \mathbf{r}_d(\frac{3L_1+1}{2}-L_2)&\dots&\mathbf{r}_d(\frac{L_1+1}{2})\end{array}\!\right],
\end{split}
\end{equation}
As desired, $L_1$ is odd, meeting the requirement of these equations\cite{cai18a}.

Based on $\mathbf{R}_u$, the UL algorithm decompose the extended covariance matrix into the following form
\begin{equation}\label{eq:Ru1}
\mathbf{R}_{u} =\mathbf{\bar A}\mathbf{R}_s\mathbf{\bar A}^H+\sigma^2\mathbf{I}
\end{equation}
where $\mathbf{\bar A}$ is further defined as
\begin{equation}
\mathbf{\bar A}=\left[\begin{array}{c}\mathbf{\bar A}_1\\ \mathbf{\bar A}_2^*\mathbf{V}^*\end{array}\right]
\end{equation}
with
\begin{equation}
\begin{split}
&\mathbf{\bar A}_{1}=[\mathbf{a}_{1}(\theta_1),\dots, \mathbf{a}_{1}(\theta_{K})] \\
&\mathbf{\bar A}_{2}=[\mathbf{a}_{2}(\theta_1),\dots, \mathbf{a}_{2}(\theta_{K})] \\
&\mathbf{a}_{1}(\theta_k)=[1,e^{-j(2\pi dsin\theta_k/\lambda)},\dots,e^{-j(2\pi d\frac{L_1-1}{2}sin\theta_k)}/\lambda]^T\\
&\mathbf{a}_{2}(\theta_k)=[1,e^{j(2\pi dsin\theta_k/\lambda)},\dots,e^{j(2\pi d(L_2-\frac{L_1+1}{2})sin\theta_k/\lambda)}]^T\\
\end{split}
\end{equation}
The dimensions of $\mathbf{\bar A}_1$ and $\mathbf{\bar A}_2$ are $\frac{L_1+1}{2} \times K$ and $(L_2-\frac{L_1-1}{2}) \times K$.

Then, eigendecomposition is applied to the positive definite Hermitian matrix $\mathbf{R}_u$, and the $(L_2+1)\times K$ signal subspace eigenvector
$\mathbf{U}_s$ and the $(L_2+1)\times (L_2+1-K)$ noise subspace eigenvector $\mathbf{U}_n$ are obtained. For an arbitrary direction $\theta$, the steering vector $\mathbf{\hat {\bar A}}(\theta)$ corresponding to $\mathbf{\bar A}$ is defined as
\begin{equation}
     \mathbf{\hat {\bar A}}=\left[\begin{array}{c}\mathbf{a}_1(\theta)\\\mathbf{a}_2^*(\theta)\rho e^{-j\psi}\end{array}\right]
\end{equation}
where $\rho$ and $\psi$ are the possible noncircularity ratio and phase of signals corresponding to $\theta$. As $\mathbf{\hat{\bar A}}$ span the signal subspace and $\mathbf{U}_{n}$ span the noise subspace, they should satisfy
\begin{equation} \label{eq:una}
      \mathbf{U}_n^H\mathbf{\hat {\bar A}}=\mathbf{0}
\end{equation}

Divide $\mathbf{U}_n$ into $\mathbf{U}_{n1}$ and $\mathbf{U}_{n2}$ with dimensions $\frac{L_1+1}{2} \times (L_2+1-K)$ and $(L_2-\frac{L_1-1}{2}) \times (L_2+1-K)$, separately.
The DOA estimation problem is then converted into the following equation
\begin{equation}\label {eq:J}
\begin{split}
 det\{\mathbf{q}^H\mathbf{M}\mathbf{q}\}=\mathbf {0}
\end{split}
\end{equation}
where
\begin{equation}
\begin{split}
    &\mathbf{q}=\left[\begin{array}{c}1\\\rho e^{j\psi}\end{array}\right]\\
    & \mathbf{M}=\left[ \begin {array}{c} \mathbf{a}_{1}^H(\theta)\mathbf{U}_{n1}\mathbf{U}_{n1}^H \mathbf{a}_{1}(\theta)\; \mathbf{ a}_{1}^H(\theta)\mathbf{U}_{n1}\mathbf{U}_{n2}^H \mathbf{a}_{2}^*(\theta)\\  \mathbf{a}_{2}^T(\theta)\mathbf{U}_{n2}\mathbf{U}_{n1}^H \mathbf{a}_{1}(\theta)\; \mathbf{a}_{2}^T(\theta)\mathbf{U}_{n2}\mathbf{U}_{n2}^H \mathbf{a}_{2}^*(\theta) \end{array}\!\right]
\end{split}
\end{equation}
where $det\{\cdot\}$ is the determinant calculator.

As $\mathbf{q}$ is a constant vector, it can be canceled. So the cost function can be written as
\begin{equation}\label{eq:19}
        \left[ \begin {array}{c} \mathbf{a}_{1}^H(\theta)\mathbf{U}_{n1}\mathbf{U}_{n1}^H \mathbf{a}_{1}(\theta)\; \mathbf{ a}_{1}^H(\theta)\mathbf{U}_{n1}\mathbf{U}_{n2}^H \mathbf{a}_{2}^*(\theta)\\  \mathbf{a}_{2}^T(\theta)\mathbf{U}_{n2}\mathbf{U}_{n1}^H \mathbf{a}_{1}(\theta)\; \mathbf{a}_{2}^T(\theta)\mathbf{U}_{n2}\mathbf{U}_{n2}^H \mathbf{a}_{2}^*(\theta) \end{array}\!\right]=\mathbf{0}
\end{equation}
The UL algorithm uses polynomial rooting to solve this minimization problem. The maximum number of mixed signals and circular signals can be resolved is
\begin{equation}\label{eq:ULK}
\begin{split}
    1\leq \bar K \leq L_2-1,  \bar K_c\leq \frac{L_1-1}{2}
\end{split}
\end{equation}

To derive the improved MUSIC (I-MUSIC) algorithm, we form a new composition of $\mathbf{M}$ as follows
\begin{equation}
\begin{split}
      \mathbf{M}&=\mathbf{\hat {\tilde A}}^H(\theta)\left[ \begin {array}{c} \mathbf{U}_{n1}\mathbf{U}_{n1}^H \; \mathbf{U}_{n1}\mathbf{U}_{n2}^H\\  \mathbf{U}_{n2}\mathbf{U}_{n1}^H \; \mathbf{U}_{n2}\mathbf{U}_{n2}^H \end{array}\!\right]\mathbf{\hat {\tilde A}}(\theta)\\
      &=\mathbf{\hat {\tilde A}}^H(\theta)\mathbf{U}_n\mathbf{U}_n^H\mathbf{\hat {\tilde A}}(\theta)
\end{split}
\end{equation}
with
\begin{equation}
\begin{split}
      \mathbf{\hat{\tilde A}}(\theta)=\left[\begin{array}{cc}\mathbf{a}_1(\theta)&\mathbf{0}\\\mathbf{0}&\mathbf{a}_2^*(\theta)\end{array}\right]\\
\end{split}
\end{equation}

As the matrix $\mathbf{M}$ is a zero-valued $2\times 2$ matrix, the determinant of it should be equal to $0$ when the cost function in Eq. (\ref{eq:19}) equals $0$. Then the DOA estimation result can be further obtained by finding peaks of the following function
\begin{equation}\label{eq:p}
\mathbf{P}(\theta)=\frac{1}{det\{\mathbf{\hat {\tilde A}}^H(\theta)\mathbf{U}_n\mathbf{ U}_n^H\mathbf{\hat {\tilde A}}(\theta)\}}
\end{equation}
where $det\{\cdot\}$ is the determinant calculator.

As the number of columns $(L_2+1-K)$ of $\mathbf{U}_n$ should be equal to or greater than $2$ to make it rank deficient, the number of signals that can be resolved can at most be $(L_2+1)-2$. The number of circular signals is restricted by the number of columns of $\mathbf{R}_{xx}$, which is one less than its rank as $\frac{L_1+1}{2}-1$. As a result, the condition of the proposed I-MUSIC is
\begin{equation}
\begin{split}
    1\leq \tilde K \leq L_2-1,  \tilde K_c\leq \frac{L_1-1}{2}
\end{split}
\end{equation}
It can be seen that the I-MUSIC can resolve exactly the same maximum number of signals as the UL algorithm.

\section{Simulation Results}
\label{sec:simulation}
In this section, simulation results are provided using the proposed I-MUSIC algorithm, in comparison with the subspace based algorithms UL, ULP and also the compressive sensing based algorithm ECM \cite{cai18a}. Consider a six-sensor nested array, and both the first and second sub-arrays have three sensors. Then, we have $L_1=23$ and $L_2=15$, and the number of signals that can be resolved is $2\leq \tilde K\leq 14$ and $\tilde K_c\leq 12$. The full angle range is from $-90^{\circ}$ to $90^{\circ}$, and the search step size is $0.1^{\circ}$ for I-MUSIC and ECM.

There are $11$ noncircular signals ($10$ BPSK signals and one PAM signal) and three circular signals, in the range $[-44^{\circ},40^{\circ}]$ with angle spacing $7^{\circ}$, which is the maximum number of noncircular signals that can be resolved by the array in theory. The number of snapshots is $12000$ and signal-to-noise ratio (SNR) is $20dB$. The simulation results are shown in Fig. 1, where we can see that as expected the I-MUSIC has resolved the maximum number of signals successfully.

\begin{figure}
\centering
\includegraphics[width=0.47\textwidth]{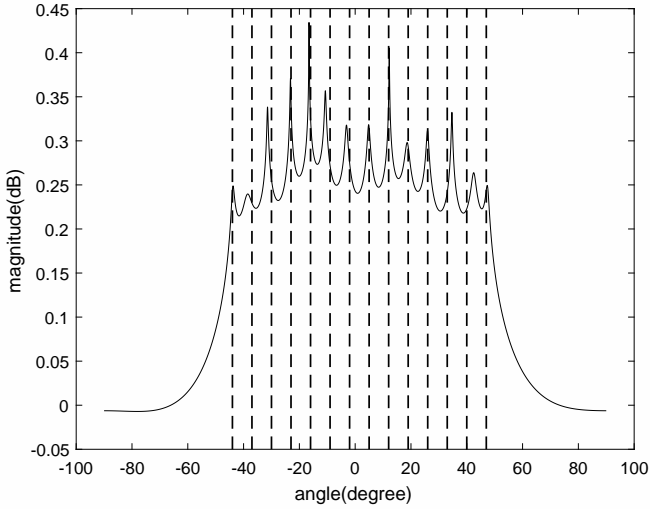}
\caption{Estimation result of 14 sources.}
\end{figure}

Next we study its estimation performance in terms of the root mean square error (RMSE). There are six signals impinging upon the array from directions $[-25^{\circ},-15^{\circ},-5^{\circ},5^{\circ},15^{\circ},25^{\circ}]$. Among them, two are circular, three are BPSK and one is PAM. The RMSE results are obtained by $500$ Monte-Carlo simulations.

First, the number of snapshots is set to be $2000$, and the SNR varies from $-10$dB to $22$dB with a step size of $4dB$. The results are shown in Fig. 3.

\begin{figure}
\centering
\includegraphics[width=0.47\textwidth]{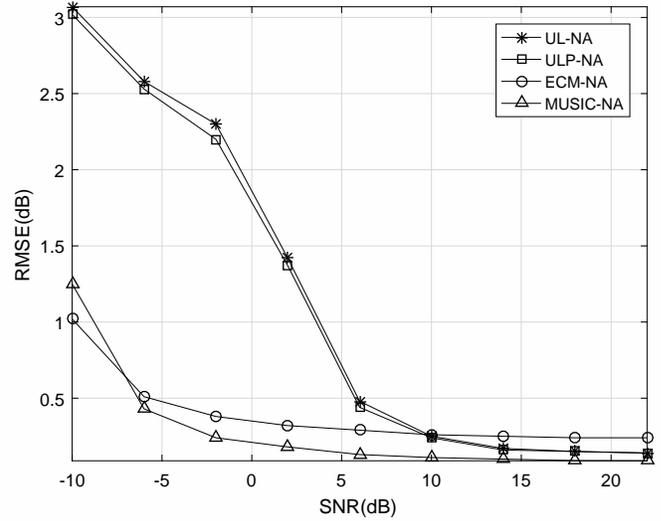}
\caption{DOA estimation results with a varying SNR.}
\label{SNRvary}
\end{figure}

It can be seen that when the SNR is equal to or greater than $-6$dB , the performance of I-MUSIC is always better than the other three algorithms. The performance of I-MUSIC is significantly better than the subspace based algorithm UL and ULP when the SNR is less than $10$dB.

\begin{figure}
\centering
\includegraphics[width=0.47\textwidth]{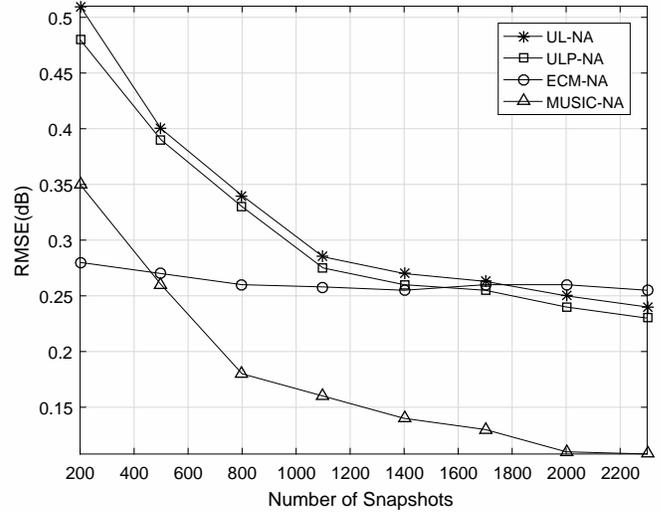}
\caption{DOA estimation results with a varying snapshot number.}
\label{Snapvary}
\end{figure}

Then, we fix the SNR at $10$ dB and change the snapshot number from $200$ to $2300$ with a step size $300$. The results are shown in Fig. \ref{Snapvary}.

As shown, when the number of snapshots is equal to or greater than $500$ , the performance of I-MUSIC is always better than the other three algorithms, especially better than UL and ULP.

\section{Conclusion}
\label{sec:conclusion}
An improved MUSIC algorithm based on the sparse array for a mixture of circular and noncircular signals has been proposed in this paper. It uses all the consecutive elements of the covariance matrix and pseudo covariance matrix to construct an extended covariance matrix, and then a MUSIC-type estimator is derived based on a new formulation of the signals. The number of resolvable signals by the proposed algorithm is $1\leq \tilde K<L_2-1$ and $\tilde K_c<\frac{L_1+1}{2}$, which is exactly the same as the previously proposed subspace based algorithm. As demonstrated by simulation results, a better performance has been achieved by the proposed I-MUSIC algorithm.

\bibliographystyle{IEEE}
\bibliography{mybib}
\end{document}